\documentclass[manuscript]{aastex}

\begin{document}
\pagenumbering{arabic}

\title{Sizes of Galactic Globular Clusters}
 
\author{Sidney van den Bergh}
\affil{Dominion Astrophysical Observatory, Herzberg Institute of Astrophysics, National Research Council of Canada, 5071 West Saanich Road, Victoria, BC, V9E 2E7, Canada}
\email{sidney.vandenbergh@nrc-cnrc.gc.ca}

\begin{abstract}  

A study is made of deviations from the mean power-law relationship between the 
Galactocentric distances and the half-light radii of Galactic globular clusters.  
 Surprisingly deviations from the mean  $R_h$ versus $R_{gc}$  relationship do not 
appear to correlate with cluster luminosity, cluster metallicity, or horizontal branch  
morphology.  Differences in orbit shape are found to contribute to the scatter
in the $R_h$ versus $R_{gc}$  relationship of Galactic globular clusters.
\end{abstract}

{\it \underline {Online material}}: extended table 
 
{\it keywords} {(Galaxy: ) Globular clusters: half-light radii}
 
 \section{INTRODUCTION}

 In a recent paper Hammer et al. (2011)  have presented persuasive evidence to
suggests that the Milky Way galaxy is a very unusual spiral galaxy, which has managed 
to avoid any major mergers since it originally formed. According to Hammer et al. 
only $\sim$ 1\% of all major spirals belong to this pristine class of galaxies. It is therefore 
of particular interest to study the group characteristics of Galactic globular clusters, 
which may which may provide information on the earliest evolutionary phase of 
galactic evolution. Fortunately we now have available from Harris (1996) (2010 edition \footnote{http://physwww.physics.mcmaster.ca/$\sim$harris/mwgc.dat})  
an essentially complete catalog of data on Galactic globular clusters. 
 
 Among the properties listed in this catalog (see Table 1) are the luminosities, metallicities,  
Galactocentric distances and half-light radii of  individual clusters. The latter parameter is  
of particular interest because it is relatively insensitive to the effects of dynamical evolution  
(Spitzer \& Thuan 1972, Lightman \& Shapiro 1978, Murphy et al. 1990). It has been shown 
 (van den Bergh 1994, 2011) that the half-light radii of Galactic globular clusters scatter  
widely around a relation of the form  
 
 ~~~~~~~~~~~~$R_h~ \alpha~ R^{2/3}_{gc}$.~~~~~~~~~~~~~~~~~~~~~~~~~~~~~~~~~~~~~~~~~~~~~~~~~~~~~~~~~~~~~~~~~~~~~~~~~~~~~~~~~(1)    
 
It is the purpose of the present investigation to ask if this scatter correlates with the   
luminosity (mass),  the metallicity [Fe/H]  or  horizontal branch characteristics of 
individual clusters.
 
 It is convenient to express the deviations of individual globular clusters from Eqn. (1) 
by a parameter that, we shall call D, which is defined by the relation 
 
~~~~~~~~~~~~~~~ D = 2/3 log $R_{gc}$  - log $R_h$.~~~~~~~~~~~~~~~~~~~~~~~~~~~~~~~~~~~~~~~~~~~~~~~~~~~~~~~~~~~~~~~~~(2) 
 
Physically a positive value of D implies that the Galactocentric distance of a cluster 
is larger than expected for its size. Alternatively one might say that clusters with  
positive D values are smaller than expected from their observed $R_{gc}$ values. 

\section{DEVIATIONS CORRELATED WITH LUMINOSITY}
 
 It was first shown by van den Bergh et al. (1991) that the radii of globular clusters 
are uncorrelated with their luminosities. This conclusion was recently strengthened 
and confirmed by van den Bergh (2011) who found that the half-light radii of Galactic  
globular clusters are independent of their  absolute magnitudes.  Figure 1 shows that the   
parameter D is also uncorrelated with cluster luminosity.  In other words, the relation  
between the half-light radii of clusters and  their Galactocentric distances also appear  
to be independent of cluster luminosity, and hence presumably of cluster mass.   
 
\section{DEVIATIONS CORRELATED WITH METALLICITY}
                                                                                                                                               
Figure 2 shows a plot of the parameter D as a function of cluster metallicity .  
Inspection of the figure hints at the possibility that metal-rich clusters with [Fe/H] $>$  
-1.0 might have systematically smaller D values than more metal-poor clusters.  
However, a Kolmogorov-Smirnov test shows that this effect falls well below any 
respectable level of statistical significance. In other words D and [Fe/H] appear  
to be uncorrelated.. Previously it had also been found (van den Bergh 2011) that  
[Fe/H] of globular clusters is independent of cluster luminosity.  
 
 \section{CORRELATIONS WITH HORIZONTAL BRANCH GRADIENT}

 Following Lee (1990) the horizontal branches of globular clusters may 
be described via the parameter C defined by the relation 
 
~~~~~~~~~~~~~~~ C = (B - R) / (B + V +R), ~~~~~~~~~~~~~~~~~~~~~~~~~~~~~~~~~~~~~~~~~~~~~~~~~~~~~~~~~~~~~~~~~(3) 
 
\noindent in which B, V and R are the number of blue, variable and red horizontal-branch 
stars, respectively. The individual values of C for halo clusters, that are given in  
Table 1, were drawn  from the compilation by MacKey \& van den Bergh (2005). 
Inspection of the data listed in Table 1, which are plotted in Figure 3,  shows not even a hint of a correlation
between the parameter D and the globular cluster horizontal branch population 
as described by the parameter C.

\section{DISCUSSION}
 
Using the recent compilation of data on Galactic globular clusters by 
Harris it is found that the half-light radii of globular clusters scale as the 
$\sim$2/3 power of their Galactocentric distances. However, the scatter about 
this relationship is considerable. In the present paper it is found that these 
deviations do not appear to correlate with either cluster luminosity, cluster 
metallicity, or population gradients along the cluster horizontal branch.  However, 
van den Bergh (1993) was able to show that the radii of individual globular  
clusters depend on orbit shape, with globulars on nearly circular orbits having  
above-average half-light radii. On the other hand clusters on retrograde orbits were 
found to have slightly below-average radii. Some speculations about the reasons 
for this dependence of size on orbit shape were given in van den Bergh (1994), but 
no firm conclusions can yet be drawn. In any case it is clear that such a 
dependence of cluster radius on orbit shape will contribute to the observed 
scatter around Equation 1.

I am indebted to Brenda Parrish and Jason Shrivell for technical support.   Special thanks are due to a particularly helpful referee.

\begin{deluxetable}{lrrrrrrcr}
\tablecolumns{9}
\tablewidth{0pt}
\tablecaption{Data on Galactic globular clusters}
\startdata
\tableline
\tableline
 
ID    &   $R_{gc}$   &   log $R_{gc}$    &   [Fe/H]  &    $M_v$   &    $R_h$  &    log $R_h$  &   D   &    C  \\

\tableline
 
N 104     & 7.4   &  0.87    &  -0.72   &   -9.42 &    4.15  &   0.62  &    -0.04  &   -0.99 \\
N 288     & 12.0  &  1.08   &   -1.32   & -6.75   &   5.77  &    0.76  &   -0.04   &  +0.98 \\
N 362     & 9.4   &  0.97    &  -1.26    & -8.43   &    2.05  &   0.31  &  +0.34  &   -0.87\\
Whi 1     & 34.5  &  1.54   &  -0.70    & -2.46   &   1.93   &   0.29   & +0.74  &    ...\\
N1261    &18.1  &  1.26   & -1.27    &  -7.80   &   3.22    &  0.51  &  +0.33   &    -0.71\\
Pal 1      & 17.2  &  1.24   & -0.65   &  -2.52    &  1.49    &  0.17   & +0.66   &   -1.00 \\
AM 1      & 124.6 &  2.10  &  -1.70   &  -4.73  &   14.71  &  1.17   & +0.23   &    -0.93 \\
Eri         &  95.0  &  1.98   & -1.43    & -5.13   &  12.06   &  1.08   & +0.24   &   -1.00 \\
Pal 2      &  35.0  &   1.54  & -1.42    & -7.97   &   3.96   &  0.60    & +0.43   &  -0.10 \\
N1851   & 16.6   &   1.22   & -1.18   & -8.33  &    1.80   &   0.26   & +0.55   &  -0.32\\
N1904   & 18.8   &  1.27   &  -1.60   & -7.86   &   2.44   &  0.39   & +0.46    &  +0.89 \\
N2298   & 15.8   & 1.20    &  -1.92   & -6.31  &   3.08   &  0.49    & +0.31   &   +0.93 \\
N2419   & 89.9   & 1.95    &  -2.15   & -9.42   &  21.38  &  1.33   &  -0.03   &    +0.86 \\
Ko 2      & 41.9   &  1.62   &  ...        &  -0.35   &   2.12  &  0.33  &  +0.75   &   ...\\
Pyx       & 41.4    & 1.62   &  -1.20   &  -5.73   &   ...      &  ...      &   ...       &     -1.00 \\
N2808   & 11.1   & 1.05   & -1.14     & -9.39    &  2.23   &  0.35   & +0.35   &   -0.49 \\
E 3        & 9.1    & 0.96   &  -0.83    &  -4.12    &   4.95  &   0.69  &   -0.05  &      ... \\
Pal 3      & 95.7  &   1.98  &  -1.63   &  -5.69   &   17.49 &   1.24 &   +0.08  &    -0.50 \\
N3201    &  8.8  &  0.94   &  -1.59    & -7.45    &   4.42   &  0.65  &  -0.02   &   +0.08\\
Pal 4      & 111.2 &  2.05  &  -1.41   &  -3.11    &   16.13 &  1.21   & +0.16  &    -1.00\\
Ko 1       &  49.3  & 1.69  &   ...       &  -4.25    &   3.65    &   0.56  & +0.57  &    ...\\
N4147    &  21.4  &  1.33 &  -1.80   &   -6.17   &    2.69  &   0.43   & +0.46  &   +0.66\\
N4372    &   7.1   &  0.85  & -2.17   &  -7.79    &  6.60    &   0.82  &  -0.25    &   +1.00  \\
Ru 106   & 18.5   &   1.27 &   -1.68  &  -6.35   &   6.48   &  0.81  &  +0.04    &  -0.82 \\
N4590     & 10.2  & 1.01   & -2.23    & -7.37    &  4.52    &   0.66  & +0.01    &  +0.17 \\
N4833     &  7.0   &   0.85  & -1.85    & -8.17    &  4.63  &   0.67  &  -0.10   &    +0.93\\
N5024    & 18.4   & 1.26    & -2.10    &  -8.71  &    6.82  &  0.83   & +0.01  &  +0.81 \\
N5053     & 17.8  &  1.25   & -2.27    &-6.76     &  13.21  &  1.12   &  -0.29   &  +0.52\\
N5139    &  6.4   & 0.81    &-1.53      &  -10.26  &   7.56  &   0.88  &  -0.34   &   ...\\
N5272    & 12.0  &  1.08   &  -1.50    &  -8.88    & 6.85    &  0.84   & -0.12    &   +0.08\\
N5286    &  8.9   &  0.95   & -1.69    &  -8.74     & 2.48    &  0.39   & +0.24   &       +0.80\\
AM 4      & 27.8   & 1.44   & -1.30    &  -1.81    &  4.03    &   0.61   &  +0.35  &   ...\\
N5466    & 16.3   & 1.21  &  -1.98    &  -6.98    & 10.70  &   1.03  &  -0.22     &   +0.58\\
N5634    &  21.2  &  1.33  &  -1.88   &  -7.69    &  6.30   &   0.80  &  +0.09    &  +0.91 \\
N5694    &  29.4   & 1.47   & -1.98    & -7.83    & 10.18  &   1.01   & -0.03     &  +1.00\\
I4499     & 15.7    & 1.20   & -1.53   &  -7.32   &  9.35    &  0.97   &  -0.17     &   +0.11\\
N5824    &  25.9   &  1.41  &  -1.91   &  -8.85  &   4.20   &  0.62   & +0.14      &  +0.79\\
Pal 5      &  18.6    &1.27   & -1.41    &   -5.17  &  18.42  &  1.27  &  -0.42      &    -0.40\\
N5897    &  7.4      &  0.87 & -1.90   &   -7.23   &   7.49   &  0.87   & -0.29      &  +0.86\\
N5904    &  6.2     & 0.79   & -1.29   &  -8.81   &   3.86   &  0.59    & -0.06     &   +0.31\\
N5927     & 4.6    &  0.66  & -0.49   &  -7.81    &  2.46    &  0.39    & +0.05    &   -1.00 \\
N5946     & 5.8   &  0.76  &  -1.29   &  -7.18    &  2.74   &  0.44    &  +0.07    &  +0.69\\
BH 176    &  12.9 & 1.11  &  0.00     & -4.06    &  4.95   &  0.69   &  +0.05     &      -1.00 \\
N5986    &   4.8   & 0.68   & -1.59   &  -8.44   &  2.96    & 0.47   &  -0.02      &   +0.97\\
Lyng 7    &  4.3    &0.63   &  -1.01   &  -6.60   &  2.79   &  0.45   &  -0.03     &    -1.00 \\
Pal 14     &  71.6  &  1.85 &  -1.62    & -4.80   & 27.15    & 1.43   &  -0.20    &   -1.00\\
N6093    &  3.8     &  0.58 & -1.75    & -8.23    & 1.77   &  0.25    & +0.14    &  +0.93\\
N6121    &  5.9    &  0.77  &  -1.16    & -7.19   &  2.77    & 0.44   &  +0.07   &   -0.06 \\
N6101    &  11.2   & 1.05   & -1.98   &  -6.94  &  4.70   &  0.67    & +0.03    &  +0.84 \\
N6144     &  2.7    & 0.43  &  -1.76   &  -6.85   &   4.22  & 0.63    & -0.34    &   +1.00 \\
N6139    &   3.6   &  0.56  & -1.65     & -8.36    & 2.50    & 0.40    & -0.03    &   +0.91 \\
Ter 3      &  2.5    & 0.40    & -0.74   &  -4.82    & 2.98    & 0.47   &  -0.20    &   -1.00 \\
N6171    & 3.3    &   0.52    &   -1.02   &     -7.12     &    3.22     &   0.51     &  -0.16    & -0.73\\
1636-2    &  2.1  &    0.32   &   -1.50    &    -4.02     &    1.21     &   0.08     &  +0.13      &  -0.40\\
N6205    &  8.4   & 0.92      &  -1.53    &   -8.55       &    3.49     &  0.54      &  +0.07     & +0.97\\
N6229   &  29.8   & 1.47     & -1.47     &   -8.06      &  3.19        &  0.50      & +0.48    &     +0.24\\
N6218    &  4.5    & 0.65     & -1.37     &  -7.31       &  2.47        &  0.39      & +0.04       &   +0.97\\
FRS173   &  3.7    &   0.57   &  ...        &   -6.45      &   0.97      &  -0.01      &  +0.39      &    ... \\
N6235     & 4.2    &  0.62    &   -1.28  &   -6.29      &  3.35       &  0.53       &  -0.12         & +0.89 \\
N6254     & 4.6    & 0.66     &  -1.56   &  -7.48       & 2.50         &  0.40      & +0.04        &  +0.98\\
N6256    &  3.0    &  0.48    & -1.02    &  -7.15       &  2.58        &  0.41      & -0.09         &   -1.00\\
Pal 15    & 38.4    &  1.58    &  -2.07   &  -5.52      & 14.43       &  1.16      &  -0.11         &   +1.00\\
N6266   &  1.7     & 0.23     &  -1.18    &  -9.18     &  1.82        & 0.26       & -0.11          &  +0.32 \\
N6273   &  1.7     &  0.23    &  -1.74    & -9.13      & 3.38        &  0.53      &  -0.38          & +0.96 \\
N6284   &  7.5     &  0.88     & -1.26    &  -7.96     &   2.94      &  0.47      & +0.12          &  +0.88 \\
N6287   &  2.1    &   0.32    &  -2.10    &  -7.36     &  2.02      & 0.31        & -0.10           &  +0.98 \\
N6293   &  1.9    &  0.28     & -1.99     &  -7.78     &  2.46     &  0.39       & -0.20            & +0.90 \\
N6304   &  2.3    &  0.36    &  -0.45     &  -7.30      & 2.44     &  0.39       & +0.15          &   -1.00 \\
N6316   &  2.6    &  0.41    & -0.45     &  -8.34      & 1.97      & 0.29       & -0.02            &   -1.00 \\
N6314   &  9.6    &  0.98    &  -2.31    &  -8.21     &  2.46      & 0.39        & +0.26          &   +0.91\\
N6325   &  1.1    &  0.04    &  -1.25    & -6.96      & 1.43       & 0.16       & -0.13            &  +0.84 \\
N6333   &  1.7    &  0.23    &  -1.77    & -7.95      &  2.21     &  0.34       & -0.19           &  +0.87 \\
N6342   &  1.7    & 0.23     &  -0.55    &  -6.42    &   1.80     &  0.26      & -0.11            &   -1.00 \\
N6356   &  7.5    & 0.88     & -0.40     & -8.51      & 3.56      & 0.55      & +0.04            &   -1.00 \\
N6355   &  1.4    &0.15     & -1.37     & -8.07      &  2.36      &  0.37      & -0.27            & +0.62 \\
N6352   &  3.3    & 0.52    & -0.64    &  -6.47     &   3.34      &   0.52     & -0.17            &   -1.00\\
I1257  &  17.9    & 1.25    & -1.70     &  -6.15    &  10.18      &  1.01     &  -0.18            &  +1.00 \\
Ter 2    & 0.8     &  -0.10  &  -0.69   &   -5.88   &     3.32     &   0.52       &    -0.59         &   -1.00\\
N6366  &   5.0   & 0.70     & -0.59    & -5.74     & 2.92        &  0.47        &  0.00             &     -0.97\\
Ter 4   & 1.0     &  0.00     & -1.41    &  -4.48    &  3.87    &   0.59          &  -0.59             &    +1.00\\
HP 1    &  0.5    &  -0.30    & -1.00   &  -6.46    &  7.39     &  0.87          &  -1.07               &   +0.75 \\
N6362   &  5.1  &   0.71    &  -0.99   &  -6.95     & 4.03    & 0.61           & -0.14              &   -0.58\\
Lil 1      &  0.8   & -0.10    &  -0.33    &  -7.32    &  ...       & ...               &  ...                 &  -1.00 \\
N6380   &    3.3  &   0.52  & -0.75    &  -7.50     & 2.35    &   0.37         &  -0.02               &   -1.00 \\
Ter 1    &  1.3     &  0.11    & -1.03    & -4.41     & 7.44    &  0.87          &  -0.80                &  -1.00 \\
Ton 2  &   1.4     & 0.15    &  -0.70     & -6.17     & 3.10    &   0.49       &  -0.39               &   -1.00\\
N6388  &   3.1   &   0.49  &  -0.55     & -9.41     & 1.50      & 0.18        &  +0.15              &   -1.00 \\
N6402  &   4.0    &  0.60   &  -1.28    &  -9.10    &  3.52     & 0.55       & -0.15                  &  +0.65\\
N6401 &    2.7    & 0.43    & -1.02     & -7.90    &  5.89      & 0.77       &  -0.48                 &  +0.35\\
N6397 &    6.0     & 0.78   &  -2.02     &  -6.64   &  1.94      & 0.29       &  +0.23            &  +0.98 \\
Pal 6   &  2.2      &  0.34    &  -0.91    &  -6.79   &   2.02    & 0.31        & -0.08               &   -1.00\\
N6426  &  14.4    &  1.16   &  -2.15    &  -6.67    &  5.51    & 0.74        & +0.03              &   +0.58\\
Djo 1   &  5.7      &  0.76    & -1.51    &  -6.98    &  ...        &  ...          &    ...                   &   ...\\
Ter 5   &  1.2       &  0.08    & -0.23     &  -7.42    &  1.45   &  0.16       &  -0.11                 &  -1.00 \\
N6440  &   1.3     & 0.11    &  -0.36     & -8.75     & 1.19    & 0.08        & -0.01                 &   -1.00\\
N6441  &   3.9     & 0.59    &  -0.46     & -9.63     & 1.92    & 0.28        &  +0.11               &   -1.00 \\
Ter 6   &  1.3       &  0.11   &  -0.56     &  -7.59    &  0.87    &-0.06       &  +0.13               &   -1.00 \\
N6453   &  3.7     &  0.57   &  -1.50     & -7.22    &  1.48     & 0.07       & +0.31                 &  +0.84 \\
UKS 1   &   0.7     &  -0.15  &  -0.64     & -6.91    &  ...       &  ...          &   ...                   &  -1.00 \\
N6496  &   4.2      &  0.62   & -0.46      & -7.20    &  3.35    &  0.53      &  -0.12            &  -1.00 \\
Ter 9    & 1.1       &  0.04     & -1.05    &   -3.71    &  1.61   &  0.21      & -0.18             &  +0.25 \\
Djo 2   &  1.8   &   0.26  & -0.65    &  -7.00  &   ...        &  ...     &    ...          &  -1.00         \\
N6517  &   4.2  &   0.62   & -1.23   &   -8.25 &    1.54 &   0.19   & +0.22    &      +0.62 \\  
Ter 10   &  2.3    & 0.36   & -1.00   &   -6.35  &   2.62  &  0.42   &   -0.18       &     -1.00 \\
N6522  &    0.6   & -0.22  &  -1.34 &   -7.65  &   2.24  &  0.35    &  -0.50    &   +0.71 \\
N6535  &   3.9   &  0.59  &  -1.79  &  -4.75   &  1.68  &  0.23     & +0.16        &    +1.00\\
N6528  &    0.6    &-0.22   & -0.11 &   -6.57  &   0.87  &  -0.06   &  -0.09        &  -1.00 \\
N6539  &   3.0    &  0.48   & -0.63 &   -8.29  &   3.86 &   0.59     & -0.27         &  -1.00  \\
N6540   &  2.8  &  0.45   & -1.35 &   -6.35     & ...      &  ...         &  ...             &         +0.30\\
N6544  &   5.1   &   0.71   & -1.40 &   -6.94   &  1.06  &   0.03     & +0.44     &   +1.00 \\
N6541   &  2.1   &   0.32  &  -1.81 &   -8.52   &  2.31   &  0.36     & -0.15      &   +1.00\\
2MS 01  &  4.5   &   0.65    & ...   &  -6.11    &  1.73    & 0.24     & +0.19           &   ...\\
ESO 06 &  14.0   &   1.15  &  -1.80  &   -4.87 &   6.54  &  0.82    & -0.05            &  ...\\
N6553    &  2.2   &   0.34  &  -0.18   &  -7.77  &  1.80  & 0.26      & -0.03           &  -1.00\\
2MS 02  &  3.2    &  0.51  &  -1.08    & -4.86  &  0.78  & -0.11    &  +0.45           &  ... \\
N6558   &  1.0    &  0.00   & -1.32    & -6.44  &  4.63 &  0.67     & -0.67         &  +0.70 \\
I1276    &  3.7   &   0.57   & -0.75    & -6.67   & 3.74 &  0.57     &  -0.19            &   -1.00\\
Ter 12   &  3.4   &    0.53   & -0.50    & -4.14  &  1.05 &  0.02    &  +0.33           &   -1.00 \\
N6569   &  3.1   &   0.49   & -0.76    & -8.28   & 2.54 &  0.40     & -0.07             &   -0.82 \\
BH 261  &   1.7   &   0.23   & -1.30   &  -4.19  &  1.04  & 0.02   &  +0.13           &     ... \\       
GLI 02   & 3.0     &  0.48   & -0.33     & ...      &    ...    &    ...    &   ...              &      ...\\
N6584   &  7.0     &   0.85   & -1.50   &  -7.69  &  2.87 &   0.46    & +0.11         &      -0.15 \\
N6624   &  1.2     &   0.08   & -0.44    & -7.49  &  1.88  &  0.27   & -0.22            &   -1.00 \\
N6626   &  2.7     &  0.43   & -1.32    & -8.16   & 3.15 &  0.50    & -0.21          &      +0.90\\
N6638   &  2.2    &   0.34  &  -0.95    & -7.12   &  1.39  &   0.14   & +0.09        &     -0.30\\
N6637   &  1.7   &   0.23   & -0.64   &  -7.64   &   2.15 &  0.33   & -0.18              &   -1.00 \\
N6642   &  1.7    &  0.23  &  -1.26   & -6.66    & 1.72   &  0.24   & -0.09          &   -0.04\\
N6652   &  2.7    &  0.43  &  -0.81  & -6.66     & 1.40    &  0.15  &  +0.14           &    -1.00\\
N6656   &  4.9     &  0.69   & -1.70 &  -8.50    &  3.13   &  0.50  &  -0.04          & +0.91\\
Pal 8      & 5.5    &  0.74  &   -0.37 &   -5.51   &   2.16  &    0.33   &  +0.16     &  -1.00\\
N6681    & 2.2   &   0.34   &  -1.62   &  -7.12    &  1.86 &   0.27   &  -0.04    &   +0.96\\
GLI 01   &  4.9   &   0.69   &   ...     & -5.91     & 0.79   &  -0.10   &  +0.56       &   ...     \\
N6712   &  3.5   &   0.54   &  -1.02   &  -7.50    &  2.67 &    0.43 &  -0.07        & -0.62\\
N6715  &  18.9   &   1.28   &  -1.49   &  -9.98    &  6.32  &  0.80 &  +0.05        &    +0.54\\
N6717   &  2.4    &  0.38   &  -1.26   &  -5.66     & 1.40  & 0.15  &  +0.10      &  +0.98\\
N6723   &   2.6   &   0.41   &  -1.10   &  -7.83    &  3.87 &    0.59  &  -0.32        &  -0.08\\
N6749   &  5.0   &   0.70   &  -1.60    & -6.70   &  2.53  &  0.40   & +0.07          &  +1.00 \\
N6752   &  5.2   &   0.72   &  -1.54   &  -7.73   &  2.22  &  0.35   & +0.13          &  +1.00\\
N6760    & 4.8   &   0.68 &  -0.40   &  -7.84   &  2.73   &  0.44  &  +0.01            &  -1.00 \\
N6779    & 9.2   &   0.96   & -1.98   &  -7.41  &   3.01  & 0.48   & +0.16            &       +0.98\\
Ter 7       & 15.6 &  1.19  &  -0.32   &  -5.01  &   5.11 &   0.71   & +0.08           &  -1.00 \\
Pal 10    &  6.4    &   0.81   & -0.10   &  -5.79  &   1.70  &   0.23   & +0.31         &   -1.00\\
Arp 2     &  21.4   &  1.33   & -1.75    & -5.29  &  14.73 &   1.17   & -0.28          &  +0.53 \\
N6809   &  3.9   &   0.59   & -1.94    & -7.57   &  4.45  &  0.65   & -0.26           &  +0.87\\
Ter 8     &  19.4   &   1.29   & -2.16   &  -5.07  &  7.27 &   0.86   &  0.00            &  +1.00\\
Pal 11   &  8.2   &   0.91  &  -0.40   &  -6.92   &  5.69  &  0.76   & -0.15             &  -1.00\\
N6838  &   6.7  &  0.83   & -0.78    & -5.61   &  1.94  &  0.29  &  +0.26               &   -1.00 \\
N6864   & 14.7  &  1.17   & -1.29    & -8.57  &   2.80 &  0.45   & +0.33             &        -0.07\\
N6934   & 12.8  &  1.11  &  -1.47    & -7.45  &  3.13 &   0.50   &  +0.24           &     +0.25\\
N6981   & 12.9  &  1.11   & -1.42    & -7.04   &  4.60 &  0.66 &  +0.08            &      +0.14\\
N7006   & 38.5  &  1.59   & -1.52   &  -7.67  &   5.27  &   0.72   & +0.34            &  -0.28\\
N7078   & 10.4  &  1.02   & -2.37   &  -9.19 &    3.03 &   0.48   & +0.20            &   +0.67\\
N7089   & 10.4  &  1.02   & -1.65    & -9.03  &   3.55 &   0.55 &  +0.13         &     +0.96\\
N7099   &   7.1  &  0.85   & -2.27    & -7.45  &  2.43 &   0.39   & +0.18          &    +0.89\\
Pal 12   &  15.8 &   1.20  &  -0.85    & -4.47  &   9.51  &  0.98   & -0.18         &    -1.00\\
Pal 13   & 26.9 &   1.43   & -1.88   &  -3.76  &   2.72  &    0.43  &  +0.52       &  -0.20 \\
N7492  &  25.3 &   1.40  &  -1.78    & -5.81  &   8.80 &   0.94   &  -0.01        &  +0.81\\

\enddata
\end{deluxetable}

\begin{figure}
\plotone{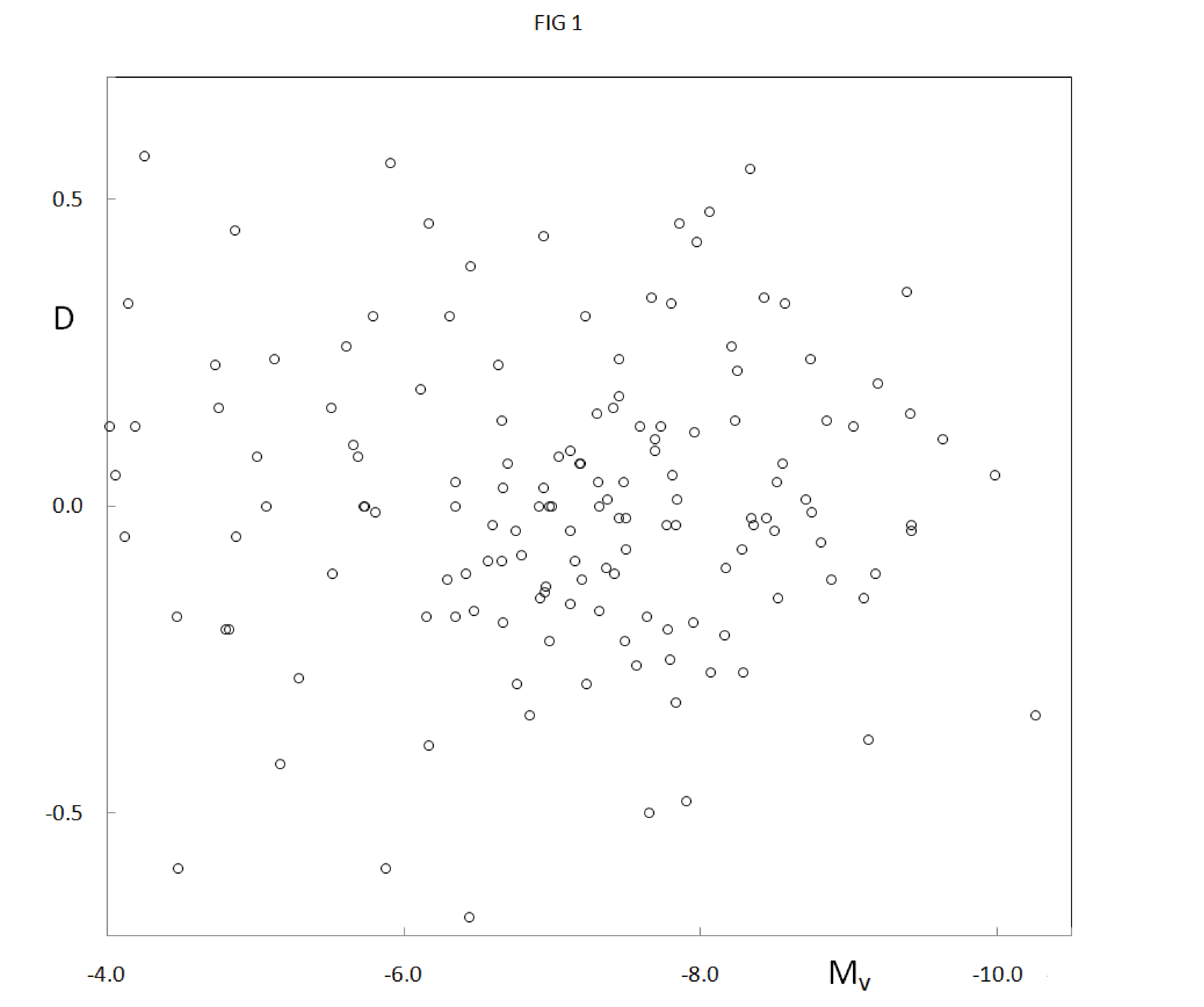}
\caption{The figure shows that the parameter D, which provides a measure of deviations from the $R_{gc}$ vs half-light radius relation of Eqn. (1), appears to be independent of cluster luminosity, and hence presumably cluster mass.}
\end{figure}
 
\begin{figure}
\plotone{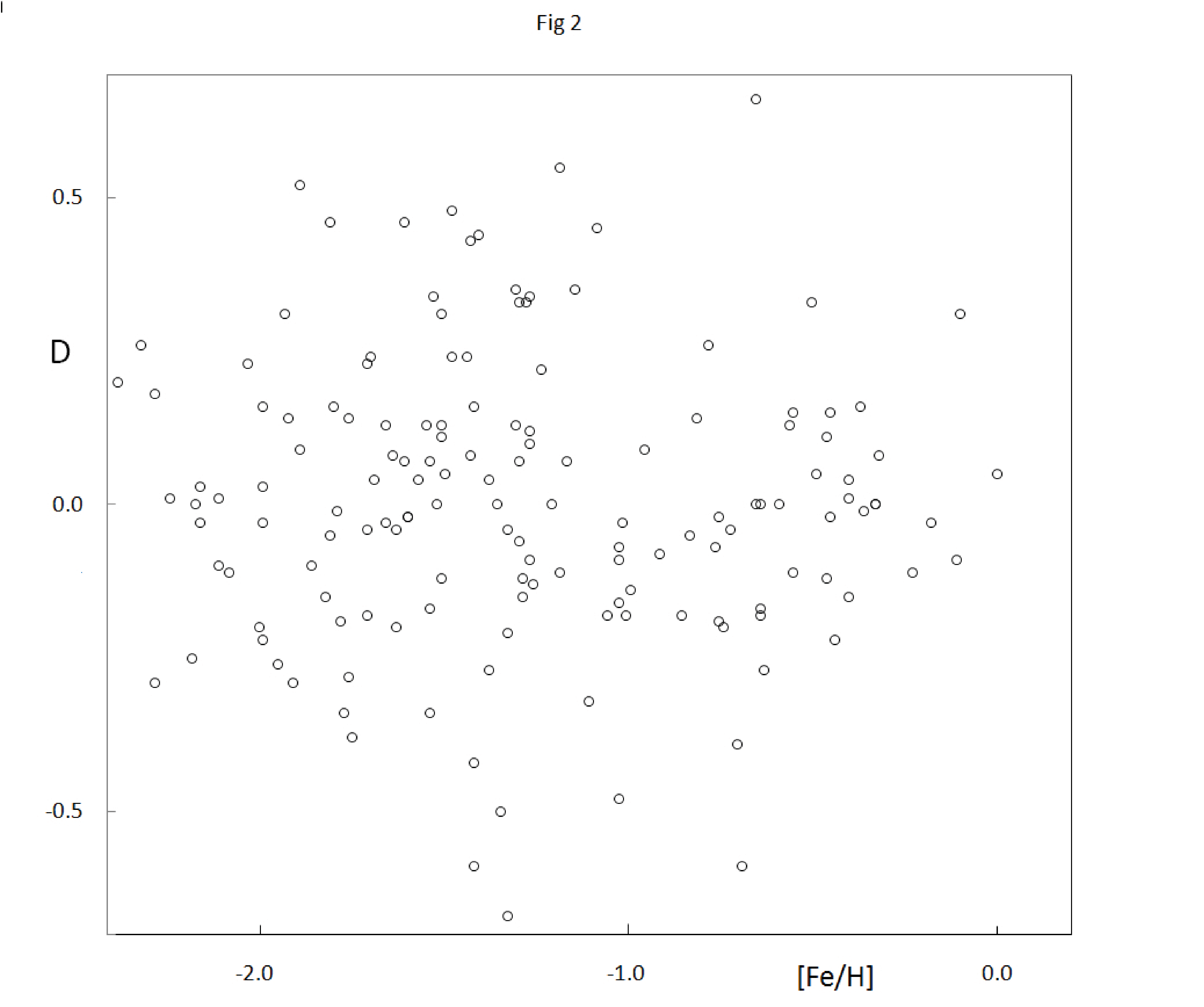}
\caption{This figure shows that the parameter D appears to be independent of the cluster metallicity [Fe/H]. In other words the deviations of clusters from the half-light radius versus Galactocentric distance relationship appear to be independent of cluster metallicity.}
\end{figure}

\begin{figure}
\plotone{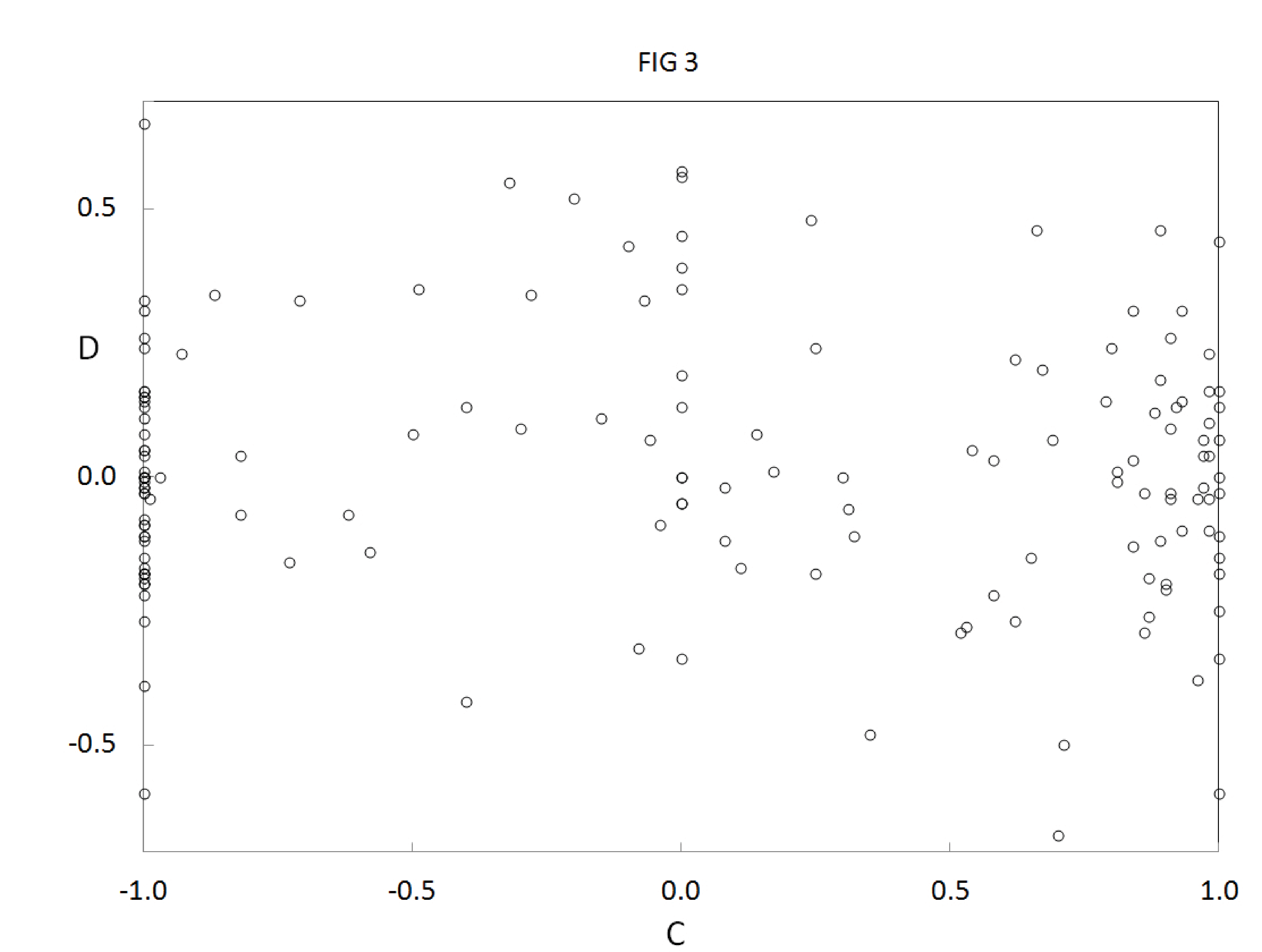}
\caption{Plot of the parameter D, that measures deviations from the $R_{gc}$ vs half-light radius relation, as a function of the horizontal branch population parameter C.  The figure shows no evidence for a correlation between C and D.}
\end{figure}


\begin{references} 

\reference{} Hammer, F., Puech, M., Flores, H., Yang, Y. B., Wang, J. L. \&  Fouquet, S. 2011, arXiv:1111.2044
\reference{} Harris, W.E. 1996, AJ, 112, 1487 
\reference{} Lee, Y.-W. 1990, ApJ, 363, 159 
\reference{} Lightman, A. P. \& Shapiro, S. L. 1978, Rev. Mod. Phys., 50, 437 
\reference{}Mackey, A. D. \& van den Bergh, S. 2005, MNRAS, 360, 631
\reference{} Murphy, B. W., Cohn, H. N. \& Hut, P. 1990, MNRAS, 245, 355 
\reference{} Spitzer, L. \& Thuan, T. X. 1972, ApJ, 175, 31 
\reference{} van den Bergh, S. 1993, ApJ, 411, 178                                  
\reference{} van den Bergh, S. 1994, AJ, 108, 2145 
\reference{} van den Bergh, S. 2011, PASP, 123, 1044 
\reference{} van den Bergh, S., Morbey, C. \& Pazder, J. 1991, ApJ, 375, 594. 
  
\end{references}
\end{document}